\documentclass[5p,times]{elsarticle}

\usepackage{amsmath}
\usepackage[mathlines]{lineno}
\usepackage{hyperref}
\usepackage{color}
\usepackage{siunitx}
\usepackage{ulem}

\usepackage{lineno}

\usepackage{multirow}

\journal{Physics Letters B}

\begin{document}

\begin{frontmatter}

\title{Beam-Normal Single Spin Asymmetry in Elastic Electron Scattering off $^{28}$Si and $^{90}$Zr}

\author[kph]{A.~Esser}\ead{esser@uni-mainz.de}
\author[kph]{M.~Thiel}
\author[kph]{P.~Achenbach}
\author[kph]{K.~Aulenbacher}
\author[kph]{S.~Aulenbacher}
\author[kph]{S.~Baunack}
\author[zagreb]{D.~Bosnar}
\author[kph]{S.~Caiazza}
\author[kph]{M.~Christmann}
\author[kph]{M.~Dehn}
\author[kph]{M.\,O.~Distler}
\author[kph]{L.~Doria}
\author[kph]{P.~Eckert}
\author[kph]{M.~Gorchtein}
\author[kph]{P.~G\"ulker}
\author[kph]{P.~Herrmann}
\author[kph]{M.~Hoek}
\author[kph]{S.~Kegel}
\author[kph]{P.~Klag}
\author[kph]{H.-J.~Kreidel}
\author[kph]{M.~Littich}
\author[kph]{S.~Lunkenheimer}
\author[kph]{F.\,E.~Maas}
\author[zagreb]{M.~Makek}
\author[kph]{H.~Merkel}
\author[kph,stefan]{M.~Mihovilovi\v{c}}
\author[kph]{J.~M\"uller}
\author[kph]{U.~M\"uller}
\author[kph]{J.~Pochodzalla}
\author[kph]{B.\,S.~Schlimme}
\author[kph]{R.~Spreckels}
\author[kph]{V.~Tioukine}
\author[kph]{C.~Sfienti}

\address[kph]{Institut f\"ur Kernphysik, Johannes Gutenberg-Universit\"at Mainz, D-55099 Mainz, Germany}
\address[zagreb]{Department of Physics, Faculty of Science, University of Zagreb, 10000 Zagreb, Croatia}
\address[stefan]{Jo\v zef Stefan Institute, SI-1000 Ljubljana, Slovenia}


\begin{abstract}
We report on a new measurement of the beam-normal single spin asymmetry $A_{\mathrm{n}}$ in the elastic scattering of 570 MeV transversely polarized electrons off $^{28}$Si and $^{90}$Zr at $Q^{2}=0.04\, \mathrm{GeV}^2/c^2$. The studied kinematics allow for a comprehensive comparison with former results on $^{12}$C.
No significant mass dependence of the beam-normal single spin asymmetry is observed in the mass regime from $^{12}$C to $^{90}$Zr.

\end{abstract}


\end{frontmatter}

\section{Beam-Normal Single Spin Asymmetry}
Parity violation in weak interactions is a well established experimental technique in atomic, particle and nuclear physics. Over the past 30 years, precision experiments have allowed to probe hadron \cite{Anthony:2005pm, Androic:2013rhu, Androic:2018kni, Maas:2004ta, Armstrong:2005hs, Aniol:2005zg} and nuclear structure \cite{Abrahamyan:2012gp} and new proposals have recently been put forward
which will considerably improve our understanding of the electroweak interaction and will allow us to explore physics beyond standard model 
\cite{Benesch:2014bas, Chen:2014psa, Becker2018}.

The interpretation of these future measurements requires theoretical predictions with uncertainties below those of the experiments. To that end it is mandatory to go beyond the one-photon exchange approximation and include higher-order corrections (such as $\gamma Z$- \cite{Gorchtein:2011mz}, $\gamma W$-,  \cite{Marciano:2005ec} or $\gamma\gamma$-box graphs \cite{Afanasev:2017gsk}) in the calculations.

The measurement of observables sensitive to two-photon exchange processes is essential to benchmark such higher-order calculations.

For this purpose the beam-normal single spin asymmetry (the so-called transverse asymmetry) $A_{\mathrm{n}}$ in polarized electron-nucleus scattering is an ideal candidate. Since $A_{\mathrm{n}}$ is a parity conserving asymmetry, arising from the interference of one- and two (or more)-photon exchange amplitudes, it gives direct access to the imaginary part of the two-photon exchange process. 

$A_{\mathrm{n}}$ can be observed when the polarization vector $\vec{P}_{\mathrm{e}}$ of the electrons is aligned parallel or antiparallel to the normal vector $\hat{n}=(\vec{k}\times \vec{k'})/|\vec{k}\times \vec{k'}|$ of the scattering plane, where $\vec{k}$ ($\vec{k'}$) are the three-momenta of the incident (scattered) electrons. The measured beam-normal single spin asymmetry in the two-photon approximation can be expressed as
\begin{linenomath*}
\begin{equation}
A_{\mathrm{n}}=\frac{\sigma^{\uparrow}-\sigma^{\downarrow}}{\sigma^{\uparrow}+\sigma^{\downarrow}}=\frac{2\,\mathrm{Im}\left(\mathcal{M}^{\ast}_{\mathrm{\gamma}}\cdot\mathcal{M}_{\mathrm{\gamma\gamma}}\right)}{\left|\mathcal{M}_{\mathrm{\gamma}}\right|^{2}},
\label{eq:An}
\end{equation}
\end{linenomath*}
where $\sigma^{\uparrow}$ ($\sigma^{\downarrow}$) denotes the cross section for electrons with spin parallel (antiparallel) to the normal vector $\hat{n}$. In Eq.~\ref{eq:An}, $\mathrm{Im}(\mathcal{M}^{\ast}_{\mathrm{\gamma}}\cdot\mathcal{M}_{\mathrm{\gamma\gamma}})$  denotes the imaginary part of the one- and two-photon exchange amplitudes $\mathcal{M}_{\gamma}$ and $\mathcal{M}_{\gamma\gamma}$ \cite{Diaconescu:2004aa}, respectively. The measured asymmetry is related to $A_{\mathrm{n}}$ by
\begin{linenomath*}
\begin{equation}
A_{\mathrm{exp}}=A_{\mathrm{n}}\vec{P}_{\mathrm{e}}\cdot\hat{n}.
\end{equation}
\end{linenomath*}
The transverse asymmetry roughly scales as $\frac{m_{\mathrm{e}}}{E}\alpha_{\mathrm{em}}$, with $m_{\mathrm{e}}$ the electron mass, $E$ the beam energy, and $\alpha_{\mathrm{em}}$ the electromagnetic coupling constant \cite{Pasquini:2005yh}. Asymmetries as small as $10^{-5}$ to $10^{-6}$ are therefore expected for beam energies of several hundred MeV. This makes the experiments particularly challenging, as statistical and systematic errors in the measurement need to be kept well below  $10^{-6}$.

The theoretical treatment of $A_\mathrm{n}$ is nontrivial as well, since the absorptive part of the two-photon exchange amplitude has to be related to the sum of all possible physical (on-mass-shell) intermediate states. 
While several approaches are available to calculate the transverse asymmetry for the reaction $p\left(e,e'\right)p$ \cite{Pasquini:2005yh,pasquini:2004,afanasev:2004,AFANASEV200448}, only two different calculations, exploiting different ansatzes, allow for extension to nuclei with $Z\geq 2$. 
Cooper and Horowitz \cite{Cooper:2005sk} are numerically solving the Dirac equation to calculate Coulomb distortion effects. To do so, they assume that only the ground state contributes, especially with increasing $Z$. In contrast, Gorchtein and Horowitz \cite{Gorchtein:2008dy} include a full range of intermediate states (elastic and inelastic), but limit their calculation to the very low four-momentum transfer region ($m_{\mathrm{e}}c\ll Q\ll E/c$).
In this model, the asymmetry can be written as:
\begin{linenomath*}
\begin{equation}
    A_\mathrm{n} \sim C_0~\mathrm{log}\left(\frac{Q^2}{m_\mathrm{e}^2c^2}\right)~\frac{F_{\mathrm{Compton}}(Q^2)}{F_{\mathrm{ch}}(Q^2)},
    \label{eq:AnTheo}
\end{equation}
\end{linenomath*}
with $C_0$ being the energy-weighted integral over the total photoabsorption cross section. It can be model-independently obtained from the optical theorem and it depends on mass number $A$ and charge number $Z$ of the target nucleus.
The last term in Eq.~\ref{eq:AnTheo}, the ratio of Compton to charge form factor, allows the model to be generalized to nuclear targets.
Up to now, measurements of the Compton slope parameter are available only for $^1$H and $^4$He targets (see \cite{Gorchtein:2008dy} and references therein).
These data suggest an approximate independence of $F_{\mathrm{Compton}}(Q^{2})/F_\mathrm{ch}(Q^{2})$ from the target nucleus.
Moreover, in the low momentum transfer region the $Q^2$ dependence of $A_\mathrm{n}$ is dominated by the logarithmic term.

\section{Previous Studies}
So far, the transverse asymmetry at forward angles ($\theta <6^{\circ}$) has been measured at the Thomas Jefferson National Accelerator Facility (JLab) for $^{1}$H, $^{4}$He, $^{12}$C, and $^{208}$Pb \cite{Abrahamyan:2012cg}. Although the data span the entire nuclear chart, a systematic interpretation in terms of $Q^{2}$, $Z$, and $E$ dependence is hindered by the different kinematics of each measurement. 
A comparison to available theoretical calculations \cite{afanasev:2004,AFANASEV200448,Gorchtein:2008dy} shows a good agreement for light nuclei, with the corresponding asymmetry being dominated by inelastic contributions.     At the same time, a striking disagreement in the case of $^{208}$Pb was observed: this may indicate the inadequacy of the two-photon exchange (TPE) approximation in \cite{Gorchtein:2008dy} given that the expansion parameter of the perturbation theory is not small ($Z\alpha\sim1$). If this were the case, the breakdown of the TPE model of Ref. \cite{Gorchtein:2008dy} should already become noticeable with intermediate heavy nuclei, thus calling for a systematic study in this mass range.

As a first step, the $Q^{2}$ dependence of $A_\mathrm{n}$ for carbon has been measured in the range between $0.02\,\mathrm{GeV}^2/c^2\ \mathrm{and}\ 0.05\,\mathrm{GeV}^2/c^2$. The obtained results show reasonable agreement with the existing theoretical calculation \cite{Esser:2018vdp}. 
The deviations from the theoretical description have been related to the assumption of the dominance of the $\log(Q^{2}/m_\mathrm{e}^{2}c^{2})$ term and the independence of $F_\mathrm{Compton}(Q^{2}) / F_\mathrm{ch}(Q^{2})$ from the target nucleus.
The result emphasizes that the $Q^{2}$ behavior of the asymmetry cannot be treat\-ed independently of the target nucleus. Even larger discrepancies could be expected for heavier nuclei.

Therefore a new experiment has been performed with the same setup and within the same four-momentum transfer range with the aim of investigating heavier target materials such as $^{28}$Si and $^{90}$Zr.

\section{New Measurements}
These experiments were carried out at the Mainz Microtron MAMI \cite{Herminghaus:1976mt} using the spectrometer setup of the A1 Collaboration \cite{Blomqvist:1998xn}, a well established facility  for high resolution spectroscopy in electron scattering experiments. To allow for comparison with previous results, the data were taken in the same kinematics as reported in \cite{Esser:2018vdp}. Minor adjustments due to the different target materials led to slightly different spectrometer angles and $Q^2$ values as given in Table\,\ref{tab:results}.
In order to study the transverse asymmetry $A_\mathrm{n}$, the A1 setup was slightly modified by inserting additional fused-silica Cherenkov detectors in the focal plane of the two high-resolution spectrometers $A$ and $B$. Corresponding to the different focal plane geometries of spectrometers $A$ and $B$, the size of the fused-silica bars were chosen to be $(300\times 70\times 10)\,\mathrm{mm}^{3}$ and $(100\times 70\times 10)\,\mathrm{mm}^{3}$, respectively. The fused-silica detectors were orientated at $45^{\circ}$ with respect to the direction of the elastically scattered electrons in the spectrometer. The produced Cherenkov light was collected by photomultiplier tubes (PMTs) with fused-silica windows. 

In the MAMI beam source, the primary electrons were produced by illuminating a strained GaAs/GaAsP super lattice photocathode with circularly polarized laser light \cite{Aulenbacher:1997zy, Aulenbacher:2011zz}. In order to measure the transverse asymmetry with the described spectrometer setup, the polarization vector of the emitted -- longitudinally polarized -- electrons had to be aligned vertically in order to be perpendicular to the scattering plane. In this two-step process, the longitudinal spin is first rotated to transverse orientation in the horizontal plane using a Wien filter \cite{TIOUKINE2006537}. Secondly, the polarization vector is rotated to the vertical orientation using a pair of solenoids. The polarization was verified to be solely vertical to within $1\%$ using a Mott polarimeter \cite{Steffens:1993ih} located downstream of the 3.5\,MeV injector linac and a M\o{}ller polarimeter \cite{moeller} close to the interaction point in the spectrometer hall. Details on this procedure can be found in \cite{Schlimme:2016rrp}. Measurements with both polarimeters determined the absolute degree of polarization. 
During each experimental campaign, the degree of polarization was monitored by frequent measurements with the Mott polarimeter.
The full range of variation of the absolute degree of polarization amongst the different measurements was between 78.2\,\% and  83.6\,\%.

The polarized electron beam had an energy of 570\,MeV and was impinging on a 1.17\,$\mathrm{g}/\mathrm{cm}^{2}$ ($1.11\,\mathrm{g}/\mathrm{cm}^{2}$) $^{28}$Si ($^{90}$Zr) target with an intensity of $20\,\mu \mathrm{A}$.
Both targets needed to be cooled during the measurement to avoid variation in their densities due to melting.
For this purpose a custom-made cooling frame was constructed. The targets with an active area of $10\,\mathrm{mm}\times 10\,\mathrm{mm}$ each were attached to a copper support structure, which was mounted on an outer aluminum frame. In this outer frame a mixture of water and ethanol was circulated at a stabilized temperature $T_\mathrm{circ}=0.5^{\circ}$C. To spread the heat load of the point-like beam spot, the electron beam was rastered over an area of $4\,\mathrm{mm} \times 4\,\mathrm{mm}$. 
For this purpose wobbler magnets have been used running mains-synchronized with a frequency of 2000 (2050)\,Hz in horizontal (vertical) direction.

\begin{figure}
\includegraphics[angle=0,width=\columnwidth]{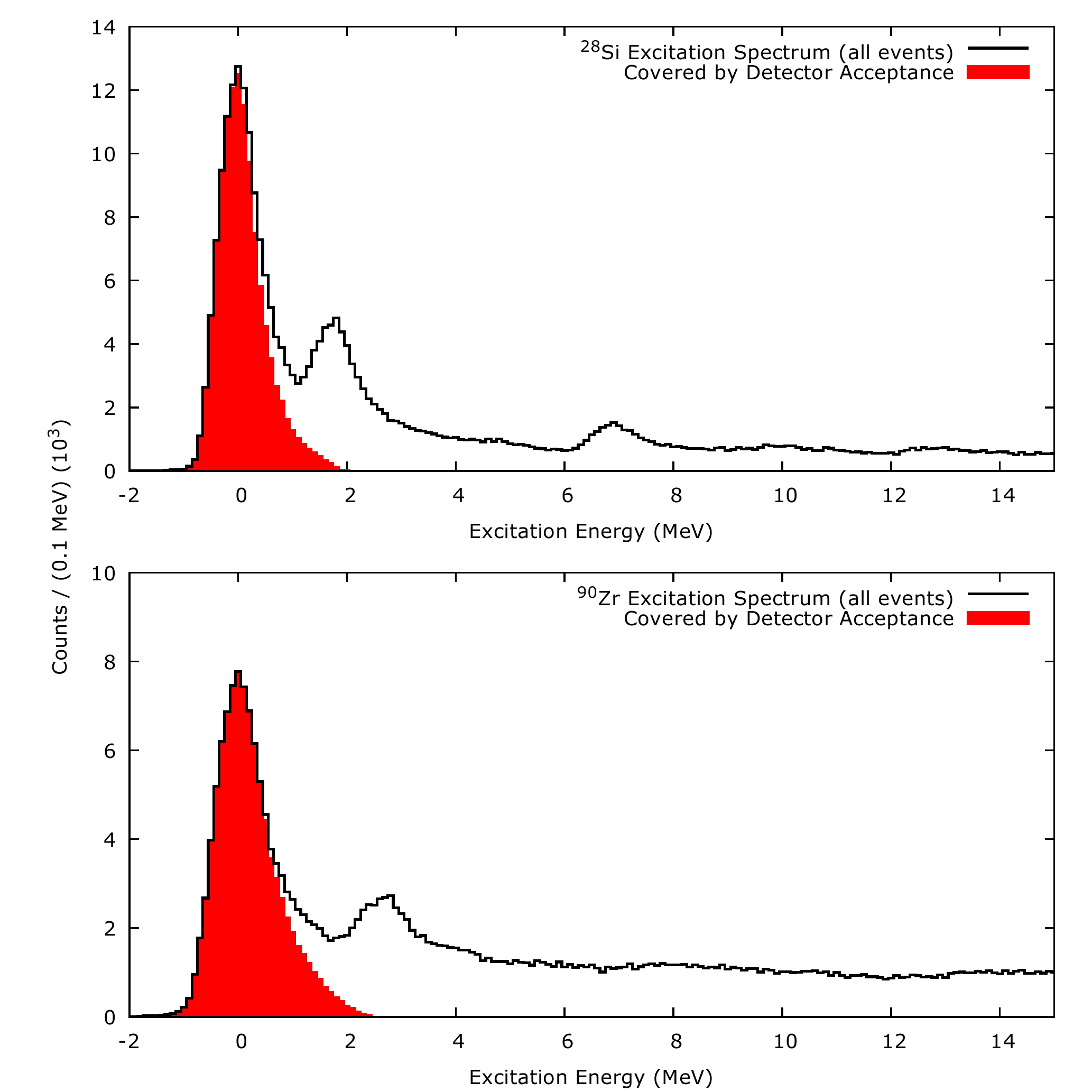}
\caption{\label{fig:exp1}The excitation energy spectra of $^{28}$Si (top panel) and $^{90}$Zr (bottom panel) show the acceptance of the spectrometer without (black line) and with (filled areas) a cut on the Cherenkov detector. By changing the magnetic field of the spectrometer the elastic peak was aligned with the position of the Cherenkov detector.}
\end{figure}

The fused-silica detectors had to be positioned such as to fully cover the elastic line while minimizing the contribution of the excited states. 
This geometrical adjustment was complicated by the small distance between the elastic peak and the first excited state of only $\Delta E \approx 1.8\,\mathrm{MeV}$ for both targets.  To verify the exact placement of the fused-silica detectors, a low beam current of $I\approx 20\,\mathrm{nA}$ was used.
In this mode, the events were processed individually by a conventional data acquisition system measuring timing and charge of the PMT pulses in parallel with the other detectors in the spectrometers.
The accurate position information obtained from a set of drift chambers allowed to match the position of the elastic line of the scattered electrons to the Cherenkov detectors by tuning the magnetic field of the high-resolution spectrometers. The resulting detector coverage is illustrated in Fig.~\ref{fig:exp1}. 
\\In the integrating mode of data taking used for the asymmetry measurements, the beam current was raised to $20\,\mu$A. 
The current produced by each detector PMT was integrated over mains-synchronized 20\,ms long periods (so-called po\-la\-ri\-za\-tion-state windows). These windows were arranged in a random sequence of quadruples with the orientation of the electron beam polarization being either $\uparrow\downarrow\downarrow\uparrow$ or $\downarrow\uparrow\uparrow\downarrow$. The polarization state was reversed by setting the high voltage of a fast Pockels cell in the optical system of the polarized electron source.
A $80\,\mu$s time window between the polarization-state windows allowed for the high voltage of the Pockels cell to be changed. The integrated PMT signal for each polarization-state window was then digitized and recorded. 

In order to identify
and reduce polarity correlated instrumental asymmetries several methods have been applied to reverse the sign of the measured asymmetry.
Besides reversing the polarization vector orientation between the measuring gates,
the differential electrical signal switching the polarity at the beam source was reversed every five minutes.
Additionally, a half-wave plate in the optical system at the beam source \cite{aulenbacher:2007} was used to reverse the beam polarity on a time scale of 24 hours. 

\begin{figure}[h]
\centering
\includegraphics[angle=0,width=0.8\columnwidth]{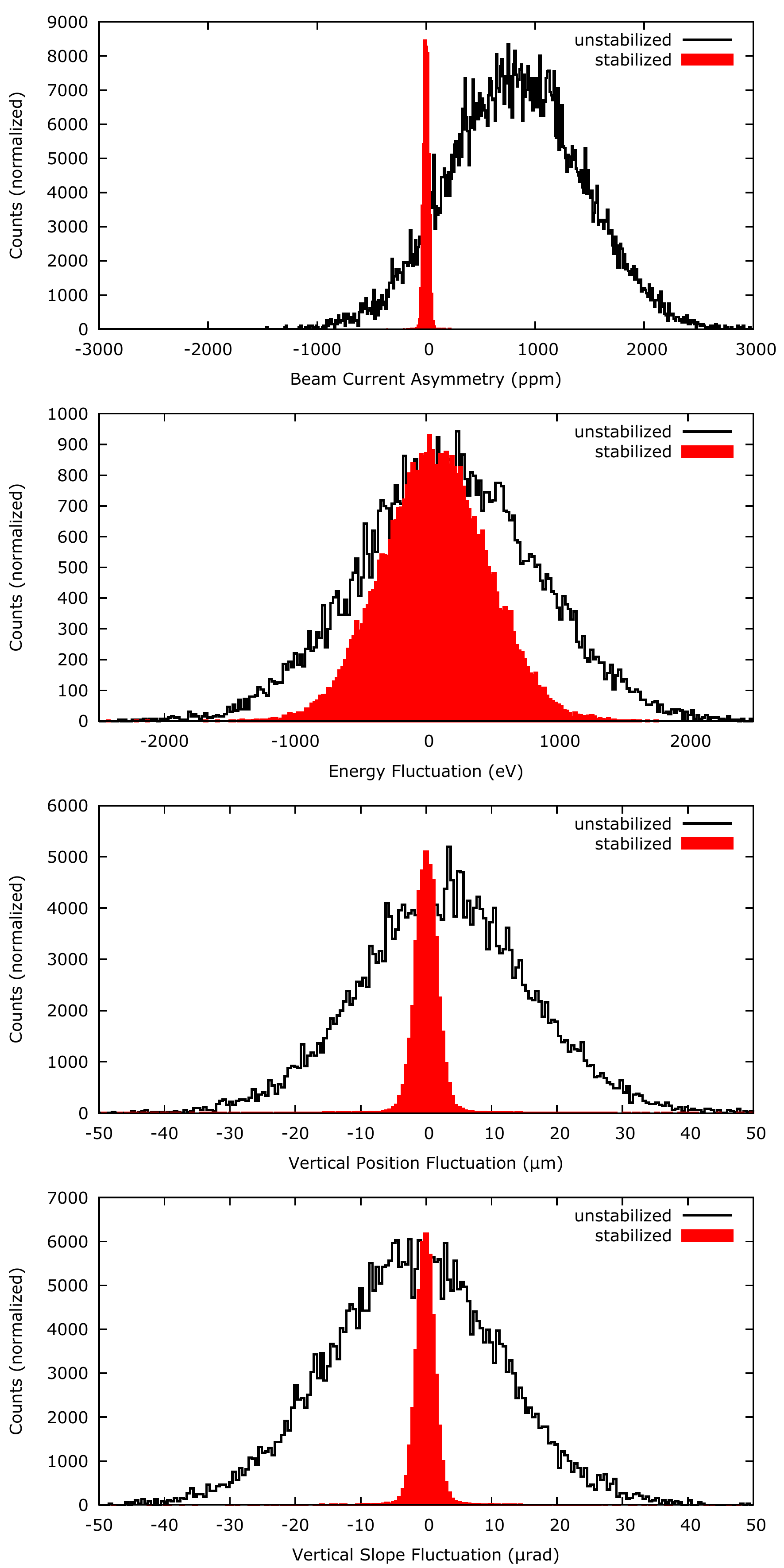}
\caption{\label{fig:exp2}Comparison between the beam parameters observed in a run with beam stabilization off (black) and with beam stabilization on (red).}
\end{figure}

Fluctuations of beam parameters such as current ($I$), energy ($E$), horizontal and vertical position ($x$ and $y$) and horizontal and vertical slope ($x'$ and $y'$) are partly correlated to the reversal of the polarization vector orientation. This can introduce instrumental asymmetries. 
Therefore it is of utmost importance to constantly control these beam parameters. They have been measured by a set of monitors, PIMO (Phase and Intensity MOnitor), ENMO (ENergy MOnitor),
and XYMO (XY MOnitor) which were used in a dedicated stabilization system to minimize polarity correlated beam fluctuations (see  Fig.~\ref{fig:exp2}) \cite{aulenbacher:2007,seidl:2000}.

In parallel, the output signals of the monitors were acquired in the same way as the detector signals,
to correct for instrumental asymmetries in the offline analysis.

\begin{table*}[!ht]
\centering
\caption{\label{tab:results}Measured beam-normal single spin asymmetries for each spectrometer and kinematical setting with the corresponding statistical and systematic uncertainty contributions in units of parts per million (ppm).} \vspace{3mm}
\begin{tabular*}{\textwidth}{clrlrlrlrl}
\hphantom{0}&Target    & \multicolumn{4}{c}{\hphantom{000000000000000}$^{28}$Si}
                       & \multicolumn{4}{c}{\hphantom{000000000000000}$^{90}$Zr} \\ \hline 
&Spectrometer          & \multicolumn{2}{c}{\hphantom{000000000000000}A}
                       & \multicolumn{2}{c}{\hphantom{0000000000000000}B}
                       & \multicolumn{2}{c}{\hphantom{0000000000000000}A}
                       & \multicolumn{2}{c}{\hphantom{0000000000000000}B}\\
&Scattering angle      & \hphantom{000000000000000}23 & .51$^{\circ}$
                       & \hphantom{0000000000000000}19 & .40$^{\circ}$
                       & \hphantom{0000000000000000}23 & .51$^{\circ}$
                       & \hphantom{0000000000000000}20 & .67$^{\circ}$ \\
&$Q^2$ (GeV$^2$/$c^2$) & 0  & .038 & 0  & .036 &  0 & .042 & 0 & .042 \\
&$A_{\mathrm{n}}$ (ppm)                & \textbf{$-$23}&\textbf{.302} &\textbf{$-$21}&\textbf{.807} & \textbf{$-$17}&\textbf{.033} & \textbf{$-$16}&\textbf{.787}\\ \hline
&\multirow{2}{*}{Total systematic error} & +0 & .553 & +0 & .386 & +1 & .390 & +1 & .527 \\
&                                        & $-$0 & .531 & $-$0 & .614 & $-$1 & .688 & $-$1 & .622 \\ \hline
&Statistical error                       & 1  & .366 & 1  & .389 & 3  & .524 & 5  & .466\\
\end{tabular*}
\end{table*}

\section{Data Analysis}
As a first step, all acquired values were corrected for fluctuations in the integration gate length.
Secondly, after the detector signals were offset-corrected, the raw asymmetry could be calculated:

\begin{linenomath*}
\begin{equation}
A_\mathrm{raw}=\frac{N_\mathrm{e}^{\uparrow}-N_\mathrm{e}^{\downarrow}}{N_\mathrm{e}^{\uparrow}+N_\mathrm{e}^{\downarrow}},
\end{equation}
\end{linenomath*}
where $N_\mathrm{e}^{\uparrow\left(\downarrow\right)}$ is the corrected detector signal.
Assuming a linear behaviour of detectors and data acquisition, $N_\mathrm{e}^{\uparrow\left(\downarrow\right)}$
is proportional to the  number of elastically scattered electrons for each polarization state.
To determine the experimental asymmetry
\begin{linenomath*}
\begin{align}
\begin{split}
A_\mathrm{exp}=A_\mathrm{raw}&-c_{1}A_\mathrm{I}-c_{2}\Delta x-c_{3}\Delta y-\\
&-c_{4}\Delta x'-c_{5}\Delta y'-c_{6}\Delta E
\label{eq:correction}
\end{split}
\end{align}
\end{linenomath*}
the raw asymmetry needs to be corrected for instrumental asymmetries.
Therefore the physical parameters of the beam have to be extracted from the beam monitor data
and the correction factors $c_{i}$ ($i=1,...,6$) need to be determined.
Due to the beam stabilization system, the helicity-correlated changes of the beam parameters were small,
but not negligible.

For the calibration of the beam monitors, dedicated runs were performed.
The PIMO signal together with the PMT gain was automatically calibrated in special runs, which have been performed approximately every three hours.
For these special runs the beam current was ramped up in steps of $0.25\,\mu$A from $17.5\,\mu$A to $22.5\,\mu$A, covering the nominal beam current setting.
The integrated PMT signal was calibrated against the beam current allowing for the extraction of an individual offset for every PMT.
This procedure also allowed to constantly check the linearity of the PMT responses and to monitor any gain variations. A precise calibration of the PIMO was also essential for the calibration of XYMOs and ENMO, since their signals scale with the beam current. 

For the XYMO calibrations, the beam was slowly rastered over a wire target with known wire positions.
For the ENMO calibration an electronic, polarity-correlated signal corresponding to a defined energy variation was superimposed on the raw energy signal.
Both, XYMOs and ENMO were calibrated once per experimental campaign.

The correction factors $c_i$ were obtained by an iterative optimization procedure.
The data were first analyzed with a given set of correction parameters. The remaining asymmetry in the detector signal was then linearly fitted against the difference of each of the beam parameters.
Finally, the correction factor was modified to counteract this effect. 
An exception to this is the correction factor $c_1$ for the beam current asymmetry, which was set to 1.

For 0.01\,\% of all events, the correction of the detector signal was not possible.
In these cases, either the gate-length signal was significantly smaller than the one chosen for the experiment
or the determined correction was larger than the physically possible value.
These events have been excluded from the analysis.

\section{Results}
The results obtained for $A_{\mathrm{n}}$ together with their uncertainties are shown in Table\,\ref{tab:results}.
Large beam fluctuations and short running time affects the $^{90}$Zr result, which exhibits larger statistical and systematic errors compared to both the $^{28}$Si measurement and our former $^{12}$C result \cite{Esser:2018vdp}.
The systematic errors consist of a set of contributions arising from different sources.
The contributions introduced by fluctuations of position, angle, and energy of the beam
were determined by varying the correction factors by $\pm\,25\,\%$, and calculating the maximum change in the resulting asymmetry. 
The same procedure was applied to the PMT signal offset allowing for a variation of up to $\pm 100\,\%$.
In a similar way, the contribution from cuts with unphysically large corrections was determined, by varying the cut threshold. 
The current and gate-length asymmetry was measured for every event.
Therefore, the remaining statistical error contributed to the systematic uncertainty.
The contribution of possible nonlinearities in the asymmetry correction was estimated by excluding 0.1\,\% of the events with the largest absolute correction for each term in Eq.\,\ref{eq:correction}, respectively. The absolute values of the resulting changes in the asymmetry were then added up.
A small difference in the number of events between the two different states of the half-wave plate and a slight variation in the measured asymmetry of both states also contributed to the systematic error.

\begin{figure}[htb]
\includegraphics[angle=0,width=\columnwidth]{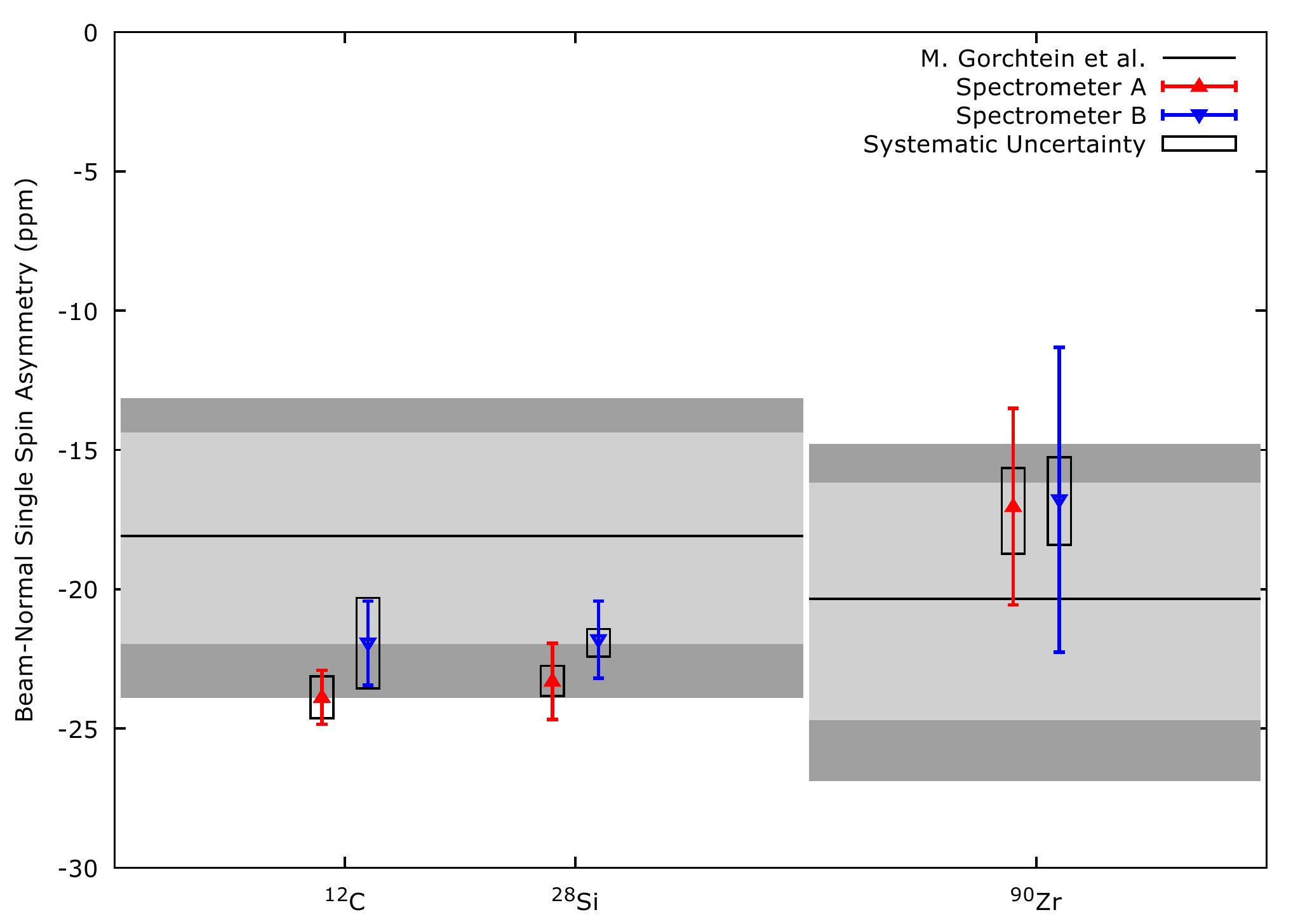}
\caption{\label{fig:result}(color online). Extracted transverse asymmetries $A_{\mathrm{n}}$ for $^{12}$C (from Ref.\,\cite{Esser:2018vdp}), $^{28}$Si and $^{90}$Zr.
The error bars mark the statistical uncertainty and the boxes show the systematic error.
The theoretical calculation for $E_{\rm b}=0.570$~GeV and  $Q^2=0.04$\,GeV/c$^2$ of Ref.~\cite{Gorchtein:2008dy} (black line) is shown for comparison.
The given bands indicate the theoretical error for uncertainties of the Compton slope parameter of \SI{10}{\%} (light grey) and \SI{20}{\%} (dark grey).}
\end{figure}

Further uncertainties in the measurement on $^{28}$Si arose from additional beam fluctuations during the XYMO calibration run.
The resulting transverse asymmetries for $^{28}$Si and $^{90}$Zr including our recent result for $^{12}$C \,\cite{Esser:2018vdp} are shown in Fig.~\ref{fig:result}
together with an extension of the theoretical calculation from Refs. \cite{Gorchtein:2008dy,Esser:2018vdp} to $^{28}$Si and $^{90}$Zr.  The error bands assigned to the theoretical predictions are computed by varying the Compton slope parameter by 10\,\% and 20\,\%.
For identical kinematics, the theoretical calculation depends only on the mass to charge ratio of the nucleus. Thus the same asymmetry is expected for both $^{12}$C and $^{28}$Si.

Within the estimated theoretical uncertainty due to the unknown Compton slope parameter of 20\,\%, the measurements are in agreement with the theoretical prediction.

A dramatic disagreement, as it was obtained for $^{208}$Pb \cite{Abrahamyan:2012cg}, has not been observed for $^{90}$Zr. Though our result is affected  by a large statistical uncertainty, its value is not compatible with zero, unlike for the $^{208}$Pb measurement. While the mean value of the asymmetry for the zirconium target slightly deviates from the values for the lighter nuclei, the experimental statistical errors and the theoretical uncertainty on the Compton slope parameter do not allow for a quantitative statement concerning a clear dependence of $A_{\mathrm{n}}$ on the nuclear charge. The discrepancy with the theoretical prediction seems to be roughly independent on the target nucleus.

The experimental results for the beam-normal single spin asymmetry on $^{28}$Si and $^{90}$Zr presented in this work 
contribute significantly to the study of this observable across the nuclear chart from hydrogen through lead. 
Our results are in agreement with all previous measurements on light and intermediate nuclei confirming that 
the theoretical model of Ref. \cite{Gorchtein:2008dy} correctly grasps the relevant physics.
Several explanations for the disagreement with the $^{208}$Pb result \cite{Abrahamyan:2012cg} are conceivable.

The coefficient $C_0$ - Eq.~\ref{eq:AnTheo} - in front of the logarithmically enhanced term could be suppressed for $^{208}$Pb.
However, this coefficient is fixed by the 
total photoabsorption cross section, which to a good approximation is known to scale with the mass of the nucleus \cite{Bianchi:1995vb}  
in the relevant energy range. Contributions from nuclear excitations are suppressed as $E_{\rm Nucl}/E_{\rm beam}$,
with $E_{\rm Nucl}$ a characteristic scale of nuclear excitations (of the order of several MeV).

A possible underestimation of the systematic uncertainty of the theoretical calculation could also explain the observed disagreement.
This uncertainty arises from two sources:
the term that is not enhanced by the large logarithm was assigned a conservative 100\,\% uncertainty; 
the Compton form factor has the exponential form, and the respective slope parameter was allowed to vary by 10\,\% - 20\,\%. 
Given the agreement of the model and the data for all nuclei up to $^{90}$Zr, an abrupt change in at least one of these terms is needed 
to reconcile the calculation with $^{208}$Pb.

Eventually  the two-photon approximation used in \cite{Gorchtein:2008dy}, while appropriate for light and intermediate mass nuclei might be inadequate for heavy nuclei.
 However, the reasonable agreement of the theory with the $^{90}$Zr data (see Fig.\,\ref{fig:result}), as well as a 
preliminary result of new calculations accounting for Coulomb distortion effects (thus summing 
corrections $\sim Z\alpha$ to all orders) \cite{koshchii:private} seem to disprove this explanation.

A new experimental program on Compton scattering at MAMI will permit to reduce the uncertainty of the Compton parameter for intermediate mass nuclei. In addition measurements of $A_\mathrm{n}$ with a $^{12}$C target at different beam energies will allow to benchmark the energy dependence of the beam-normal single spin asymmetry in the theoretical treatment.
Furthermore a new measurement of $A_\mathrm{n}$ for $^{208}$Pb by the PREX-II experiment \cite{prex2proposal} might provide additional clues to the solution of the current tension.

\section*{Acknowledgments}
We acknowledge the MAMI accelerator group and all the workshop staff members for outstanding support. This work was supported by the PRISMA+ (Precision Physics, Fundamental Interactions and Structure of Matter) Cluster of Excellence, the Deutsche Forschungsgemeinschaft through the Collaborative Research Center 1044, and the Federal State of Rhine\-land-Palatinate. The work of M. Gorchtein was supported by the German-Mexican research collaboration grant No. 278017 (CONACyT) and No. SP 778/4- 1 (DFG), and by the EU Horizon 2020 research and innovation programme, project STRONG- 2020, grant agreement No. 824093. 
%
\bibliography{TransverseSi28}

\begin{thebibliography}{10}
\expandafter\ifx\csname url\endcsname\relax
  \def\url#1{\texttt{#1}}\fi
\expandafter\ifx\csname urlprefix\endcsname\relax\def\urlprefix{URL }\fi
\expandafter\ifx\csname href\endcsname\relax
  \def\href#1#2{#2} \def\path#1{#1}\fi

\bibitem{Anthony:2005pm}
P.~L. Anthony, et~al., {Precision measurement of the weak mixing angle in
  M{\o}ller scattering}, Phys. Rev. Lett. 95 (2005) 081601.
\newblock \href {http://arxiv.org/abs/hep-ex/0504049}
  {\path{arXiv:hep-ex/0504049}}, \href
  {https://doi.org/10.1103/PhysRevLett.95.081601}
  {\path{doi:10.1103/PhysRevLett.95.081601}}.

\bibitem{Androic:2013rhu}
D.~Androic, et~al., {First Determination of the Weak Charge of the Proton},
  Phys. Rev. Lett. 111~(14) (2013) 141803.
\newblock \href {http://arxiv.org/abs/1307.5275} {\path{arXiv:1307.5275}},
  \href {https://doi.org/10.1103/PhysRevLett.111.141803}
  {\path{doi:10.1103/PhysRevLett.111.141803}}.

\bibitem{Androic:2018kni}
D.~Androic, et~al., {Precision measurement of the weak charge of the proton},
  Nature 557~(7704) (2018) 207--211.
\newblock \href {https://doi.org/10.1038/s41586-018-0096-0}
  {\path{doi:10.1038/s41586-018-0096-0}}.

\bibitem{Maas:2004ta}
F.~E. Maas, et~al., {Measurement of strange quark contributions to the
  nucleon's form-factors at Q$^2$ = 0.230 (GeV/c)$^2$}, Phys. Rev. Lett. 93
  (2004) 022002.
\newblock \href {http://arxiv.org/abs/nucl-ex/0401019}
  {\path{arXiv:nucl-ex/0401019}}, \href
  {https://doi.org/10.1103/PhysRevLett.93.022002}
  {\path{doi:10.1103/PhysRevLett.93.022002}}.

\bibitem{Armstrong:2005hs}
D.~S. Armstrong, et~al., {Strange quark contributions to parity-violating
  asymmetries in the forward G0 electron-proton scattering experiment}, Phys.
  Rev. Lett. 95 (2005) 092001.
\newblock \href {http://arxiv.org/abs/nucl-ex/0506021}
  {\path{arXiv:nucl-ex/0506021}}, \href
  {https://doi.org/10.1103/PhysRevLett.95.092001}
  {\path{doi:10.1103/PhysRevLett.95.092001}}.

\bibitem{Aniol:2005zg}
K.~A. Aniol, et~al., {Constraints on the nucleon strange form-factors at Q$^2
  \sim 0.1\,$GeV$^2$}, Phys. Lett. B635 (2006) 275--279.
\newblock \href {http://arxiv.org/abs/nucl-ex/0506011}
  {\path{arXiv:nucl-ex/0506011}}, \href
  {https://doi.org/10.1016/j.physletb.2006.03.011}
  {\path{doi:10.1016/j.physletb.2006.03.011}}.

\bibitem{Abrahamyan:2012gp}
S.~Abrahamyan, et~al., {Measurement of the Neutron Radius of $^{208}$Pb Through
  Parity-Violation in Electron Scattering}, Phys. Rev. Lett. 108 (2012) 112502.
\newblock \href {http://arxiv.org/abs/1201.2568} {\path{arXiv:1201.2568}},
  \href {https://doi.org/10.1103/PhysRevLett.108.112502}
  {\path{doi:10.1103/PhysRevLett.108.112502}}.

\bibitem{Benesch:2014bas}
J.~Benesch, et~al., {The MOLLER Experiment: An Ultra-Precise Measurement of the
  Weak Mixing Angle Using M{\o}ller Scattering} (2014).
\newblock \href {http://arxiv.org/abs/1411.4088} {\path{arXiv:1411.4088}}.

\bibitem{Chen:2014psa}
J.~Chen, et~al., {A White Paper on SoLID (Solenoidal Large Intensity Device)}
  (2014).
\newblock \href {http://arxiv.org/abs/1409.7741} {\path{arXiv:1409.7741}}.

\bibitem{Becker2018}
D.~Becker, et~al., \href{https://doi.org/10.1140/epja/i2018-12611-6}{{The P2
  Experiment}}, Eur. Phys. J. A 54~(11) (2018) 208.
\newblock \href {http://arxiv.org/abs/1802.04759} {\path{arXiv:1802.04759}},
  \href {https://doi.org/10.1140/epja/i2018-12611-6}
  {\path{doi:10.1140/epja/i2018-12611-6}}.
\newline\urlprefix\url{https://doi.org/10.1140/epja/i2018-12611-6}

\bibitem{Gorchtein:2011mz}
M.~Gorchtein, C.~J. Horowitz, M.~J. Ramsey-Musolf, {Model-dependence of the
  $\gamma Z$ dispersion correction to the parity-violating asymmetry in elastic
  $ep$ scattering}, Phys. Rev. C84 (2011) 015502.
\newblock \href {http://arxiv.org/abs/1102.3910} {\path{arXiv:1102.3910}},
  \href {https://doi.org/10.1103/PhysRevC.84.015502}
  {\path{doi:10.1103/PhysRevC.84.015502}}.

\bibitem{Marciano:2005ec}
W.~J. Marciano, A.~Sirlin, {Improved calculation of electroweak radiative
  corrections and the value of $V_\mathrm{ud}$}, Phys. Rev. Lett. 96 (2006)
  032002.
\newblock \href {http://arxiv.org/abs/hep-ph/0510099}
  {\path{arXiv:hep-ph/0510099}}, \href
  {https://doi.org/10.1103/PhysRevLett.96.032002}
  {\path{doi:10.1103/PhysRevLett.96.032002}}.

\bibitem{Afanasev:2017gsk}
A.~Afanasev, P.~G. Blunden, D.~Hasell, B.~A. Raue, {Two-photon exchange in
  elastic electron–proton scattering}, Prog. Part. Nucl. Phys. 95 (2017)
  245--278.
\newblock \href {http://arxiv.org/abs/1703.03874} {\path{arXiv:1703.03874}},
  \href {https://doi.org/10.1016/j.ppnp.2017.03.004}
  {\path{doi:10.1016/j.ppnp.2017.03.004}}.

\bibitem{Diaconescu:2004aa}
L.~Diaconescu, M.~J. Ramsey-Musolf, {The Vector analyzing power in elastic
  electron-proton scattering}, Phys. Rev. C70 (2004) 054003.
\newblock \href {http://arxiv.org/abs/nucl-th/0405044}
  {\path{arXiv:nucl-th/0405044}}, \href
  {https://doi.org/10.1103/PhysRevC.70.054003}
  {\path{doi:10.1103/PhysRevC.70.054003}}.

\bibitem{Pasquini:2005yh}
B.~Pasquini, M.~Vanderhaeghen, {Single spin asymmetries in elastic
  electron-nucleon scattering}, Eur.\ Phys.\ J.\ A 24 (2005) 29--32.
\newblock \href {http://arxiv.org/abs/hep-ph/0502144}
  {\path{arXiv:hep-ph/0502144}}, \href
  {https://doi.org/10.1140/epjad/s2005-04-005-3}
  {\path{doi:10.1140/epjad/s2005-04-005-3}}.

\bibitem{pasquini:2004}
B.~Pasquini, M.~Vanderhaeghen,
  \href{https://link.aps.org/doi/10.1103/PhysRevC.70.045206}{Resonance
  estimates for single spin asymmetries in elastic electron-nucleon
  scattering}, Phys. Rev. C 70 (2004) 045206.
\newblock \href {https://doi.org/10.1103/PhysRevC.70.045206}
  {\path{doi:10.1103/PhysRevC.70.045206}}.
\newline\urlprefix\url{https://link.aps.org/doi/10.1103/PhysRevC.70.045206}

\bibitem{afanasev:2004}
A.~V. Afanasev, N.~P. Merenkov,
  \href{https://link.aps.org/doi/10.1103/PhysRevD.70.073002}{Large logarithms
  in the beam normal spin asymmetry of elastic electron-proton scattering},
  Phys. Rev. D 70 (2004) 073002.
\newblock \href {https://doi.org/10.1103/PhysRevD.70.073002}
  {\path{doi:10.1103/PhysRevD.70.073002}}.
\newline\urlprefix\url{https://link.aps.org/doi/10.1103/PhysRevD.70.073002}

\bibitem{AFANASEV200448}
A.~V. Afanasev, N.~P. Merenkov,
  \href{http://www.sciencedirect.com/science/article/pii/S0370269304011700}{Collinear
  photon exchange in the beam normal polarization asymmetry of elastic
  electron–proton scattering}, Physics Letters B 599~(1) (2004) 48 -- 54.
\newblock \href {https://doi.org/10.1016/j.physletb.2004.08.023}
  {\path{doi:10.1016/j.physletb.2004.08.023}}.
\newline\urlprefix\url{http://www.sciencedirect.com/science/article/pii/S0370269304011700}

\bibitem{Cooper:2005sk}
E.~D. Cooper, C.~J. Horowitz, {The Vector analyzing power in elastic
  electron-nucleus scattering}, Phys. Rev. C72 (2005) 034602.
\newblock \href {http://arxiv.org/abs/nucl-th/0506034}
  {\path{arXiv:nucl-th/0506034}}, \href
  {https://doi.org/10.1103/PhysRevC.72.034602}
  {\path{doi:10.1103/PhysRevC.72.034602}}.

\bibitem{Gorchtein:2008dy}
M.~Gorchtein, C.~J. Horowitz, {Analyzing power in elastic scattering of the
  electrons off a spin-0 target}, Phys. Rev. C77 (2008) 044606.
\newblock \href {http://arxiv.org/abs/0801.4575} {\path{arXiv:0801.4575}},
  \href {https://doi.org/10.1103/PhysRevC.77.044606}
  {\path{doi:10.1103/PhysRevC.77.044606}}.

\bibitem{Abrahamyan:2012cg}
S.~Abrahamyan, et~al., {New Measurements of the Transverse Beam Asymmetry for
  Elastic Electron Scattering from Selected Nuclei}, Phys. Rev. Lett. 109
  (2012) 192501.
\newblock \href {http://arxiv.org/abs/1208.6164} {\path{arXiv:1208.6164}},
  \href {https://doi.org/10.1103/PhysRevLett.109.192501}
  {\path{doi:10.1103/PhysRevLett.109.192501}}.

\bibitem{Esser:2018vdp}
A.~Esser, et~al., {First Measurement of the $Q^2$ Dependence of the Beam-Normal
  Single Spin Asymmetry for Elastic Scattering off Carbon}, Phys. Rev. Lett.
  121~(2) (2018) 022503.
\newblock \href {https://doi.org/10.1103/PhysRevLett.121.022503}
  {\path{doi:10.1103/PhysRevLett.121.022503}}.

\bibitem{Herminghaus:1976mt}
H.~Herminghaus, A.~Feder, K.~H. Kaiser, W.~Manz, H.~Von Der~Schmitt, {The
  Design of a Cascaded 800-MeV Normal Conducting CW Racetrack Microtron}, Nucl.
  Instrum. Meth. 138 (1976) 1--12.
\newblock \href {https://doi.org/10.1016/0029-554X(76)90145-2}
  {\path{doi:10.1016/0029-554X(76)90145-2}}.

\bibitem{Blomqvist:1998xn}
K.~I. Blomqvist, et~al., {The three-spectrometer facility at the Mainz
  microtron MAMI}, Nucl. Instrum. Meth. A403 (1998) 263--301.
\newblock \href {https://doi.org/10.1016/S0168-9002(97)01133-9}
  {\path{doi:10.1016/S0168-9002(97)01133-9}}.

\bibitem{Aulenbacher:1997zy}
K.~Aulenbacher, et~al., {The MAMI source of polarized electrons}, Nucl.
  Instrum. Meth. A391 (1997) 498--506.
\newblock \href {https://doi.org/10.1016/S0168-9002(97)00528-7}
  {\path{doi:10.1016/S0168-9002(97)00528-7}}.

\bibitem{Aulenbacher:2011zz}
K.~Aulenbacher, {Polarized beams for electron accelerators}, Eur. Phys. J. ST
  198 (2011) 361--380.
\newblock \href {https://doi.org/10.1140/epjst/e2011-01499-6}
  {\path{doi:10.1140/epjst/e2011-01499-6}}.

\bibitem{TIOUKINE2006537}
V.~Tioukine, K.~Aulenbacher,
  \href{http://www.sciencedirect.com/science/article/pii/S0168900206014434}{{Operation
  of the MAMI accelerator with a Wien filter based spin rotation system}},
  Nucl. Instrum. Meth. A 568~(2) (2006) 537 -- 542.
\newblock \href {https://doi.org/https://doi.org/10.1016/j.nima.2006.08.022}
  {\path{doi:https://doi.org/10.1016/j.nima.2006.08.022}}.
\newline\urlprefix\url{http://www.sciencedirect.com/science/article/pii/S0168900206014434}

\bibitem{Steffens:1993ih}
K.~H. Steffens, H.~G. Andresen, J.~Blume-Werry, F.~Klein, K.~Aulenbacher,
  E.~Reichert, {A Spin rotator for producing a longitudinally polarized
  electron beam with MAMI}, Nucl. Instrum. Meth. A325 (1993) 378--383.
\newblock \href {https://doi.org/10.1016/0168-9002(93)90383-S}
  {\path{doi:10.1016/0168-9002(93)90383-S}}.

\bibitem{moeller}
A.~Tyukin, Master Thesis, Inst. f. Kernphysik, Johannes Gutenberg-Universit\"at
  Mainz (2015).

\bibitem{Schlimme:2016rrp}
B.~S. Schlimme, et~al., {Vertical Beam Polarization at MAMI}, Nucl. Instrum.
  Meth. A850 (2017) 54--60.
\newblock \href {http://arxiv.org/abs/1612.02863} {\path{arXiv:1612.02863}},
  \href {https://doi.org/10.1016/j.nima.2017.01.024}
  {\path{doi:10.1016/j.nima.2017.01.024}}.

\bibitem{aulenbacher:2007}
K.~Aulenbacher, Helicity correlated asymmetries caused by optical
  imperfections, Eur. Phys. J. A 32 (2007) 543 -- 547.
\newblock \href {https://doi.org/10.1140/epja/i2006-10397-8}
  {\path{doi:10.1140/epja/i2006-10397-8}}.

\bibitem{seidl:2000}
M.~Seidl, et~al.,
  \href{http://accelconf.web.cern.ch/e00/PAPERS/WEP3B18.pdf}{{High Precision
  Beam Energy Stabilisation of the Mainz Microtron MAMI}}, in: Proceedings of
  the 7th European Particle Accelerator Conference, 2000, p. 1930.
\newline\urlprefix\url{http://accelconf.web.cern.ch/e00/PAPERS/WEP3B18.pdf}

\bibitem{Bianchi:1995vb}
N.~Bianchi, et~al.,
  \href{https://link.aps.org/doi/10.1103/PhysRevC.54.1688}{Total hadronic
  photoabsorption cross section on nuclei in the nucleon resonance region},
  Phys. Rev. C 54 (1996) 1688--1699.
\newblock \href {https://doi.org/10.1103/PhysRevC.54.1688}
  {\path{doi:10.1103/PhysRevC.54.1688}}.
\newline\urlprefix\url{https://link.aps.org/doi/10.1103/PhysRevC.54.1688}

\bibitem{koshchii:private}
O.~Koshchii, private communication.

\bibitem{prex2proposal}
K.~Paschke, et~al.,
  \href{https://hallaweb.jlab.org/parity/prex/prexII.pdf}{{Proposal to
  Jefferson Lab PAC 38 - Prex-II: Precision Parity-Violating Measurement of the
  Neutron Skin of Lead}}.
\newline\urlprefix\url{https://hallaweb.jlab.org/parity/prex/prexII.pdf}

\end{thebibliography}

\end{document}